\documentclass[12pt]{article}
\usepackage{amssymb}

\newtheorem{Prop}{Proposition}
\newcommand{\R}{{\mathbb R}}  
\newcommand{\Rplus}{\R_+}

\newtheorem{Teo}{Theorem}
\newtheorem{Lema}{Lemma}

\newtheorem{Def}{Definition}

\newcommand{\Dyn}[0]{\Phi}

\begin{document}

\date{}

\title{On the Stability of Murray's Testosterone Model}
\author{German Enciso%
\footnote{to whom correspondence should be addressed.  E-mail: {\tt enciso@math.rutgers.edu}.  
Department of Mathematics,
Hill Center-Busch Campus,
Rutgers, The State University of New Jersey,
110 Frelinghuysen Rd,
Piscataway, NJ 08854-8019, USA, tel. (01) 732 445 2390.  Supported in part by AFOSR Grant F49620-01-1-0063,
NIH Grant P20 GM64375, and Dimacs.}%
\quad Eduardo D. Sontag%
\footnote{E-mail: {\tt sontag@control.rutgers.edu}.
Supported in part by AFOSR Grant F49620-01-1-0063 and
NIH Grant R01 GM46383.}%
\\
Department of Mathematics\\
Rutgers University\\
New Brunswick, NJ 08903}

\maketitle

\begin{abstract}

We prove the global asymptotic stability of a well-known
delayed negative-feedback model
of testosterone dynamics, which has been proposed as a model of oscillatory
behavior. 
We establish stability (and hence the impossibility of oscillations) even in
the presence of delays of arbitrary length. 

\end{abstract}


\section{Introduction}

The concentration of testosterone in the blood of a healthy human male is
known to oscillate periodically every few hours, in response to similar
oscillations in the concentrations of the luteinising hormone (LH) secreted by
the pituitary gland, and the luteinising hormone releasing hormone (LHRH),
normally secreted by the hypothalamus (see~\cite{Cart 1986},\cite{Smith 1980}).
In his influential textbook {\em Mathematical Biology}~\cite{Murray 2002},
J.D. Murray presents this process as an example of a biological oscillator,
and proposes a model to describe it (pp. 244-253 in this edition).  To obtain
oscillations in an otherwise stable model, he introduces a delay in one of the
variables, and by linearizing around the unique equilibrium point, he presents
an argument to find conditions for the existence of such oscillations.  This
section in his book has remained virtually unchanged since the first edition
of 1989, up to the recent publication of the third edition in 2002.

The study of delayed models is one of great interest for its applicability in
biological applications since it introduces a very relevant realism.
(Consider for instance the delay between the moment a protein is transcribed,
and the moment the folded and translated protein gets to act as a
transcription factor back in the nucleus.)   
This realism often comes at
the expense of a higher difficulty in mathematical treatment.

As a ``case study'' for a method for proving stability in
a class of dynamical systems with delays, we show in this paper that 
{\em Murray's model in fact does not exhibit oscillations.}
The biological model itself, while simplified, is still interesting in its own
right, and belongs to a commonly recurring class of models of negative
feedback proposed (in undelayed form) by Goodwin~\cite{Goodwin 1965}, and
illustrated in Goldbeter~\cite{Goldbeter 1996}.  In what follows, we first
study the linearized system around the unique equilibrium, establishing local
stability, and 
then proceed to show the global stability of the system, borrowing ideas from
monotone systems and the theory of control.
We also propose an explanation for the confusion in~\cite{Murray 2002}.

\section{The Model, And Its Linearization}

The presence of LHRH in the blood is assumed in this simple model to induce
the secretion of LH, which induces testosterone to be secreted in the testes.
The testosterone in turn causes a negative feedback effect on the secretion of
LHRH.  Denoting LHRH, LH, and testosterone by $R,L$, and $T$ respectively, and
assuming first order degradation and a delay $\tau$ in the response of the
testes to changes in LH, we arrive to the dynamical system 
\begin{equation} 
      \begin{array}{l} 
         \dot{R}=f(T)-b_1 R \\
         \dot{L}=g_1 R - b_2 L \\
         \dot{T}=g_2 L(t-\tau) - b_3 T 
      \end{array}
\label{eqmodel}
\end{equation}
Here, $b_1,b_2,b_3,g_1,g_2$ are positive constants, $\tau\geq 0$ and
$f(x)=A/(K+x)$, although other positive, monotone decreasing functions could
be employed as well (see Murray, p. 246).   

By setting the left hand sides equal to zero, it is straightforward to show
that there are as many equilibrium points of (\ref{eqmodel}) as there are
solutions of
\begin{equation}
f(T)-\frac{b_1b_2b_3T}{g_1g_2}=0 \label{eqT}
\end{equation}
namely for each such solution $T_0$ of (\ref{eqT}), one has the equilibrium 
\begin{equation}
L_0=\frac{b_3T_0}{g_2}, \ R_0=\frac{b_3b_2T_0}{g_1g_2},\ T_0
\end{equation}
and by the assumption of positivity and monotonicity of $f$ there always exists
a unique solution of (\ref{eqT}), thus a unique equilibrium point of
(\ref{eqmodel}).  Linearizing around that point we obtain the system  

\begin{equation}
     \begin{array}{l}
        \dot{x}=f'(T_0)z - b_1 x  \\
        \dot{y}=g_1 x - b_2 y     \\
        \dot{z}=g_2 y(t-\tau) - b_3 z
     \end{array}
\label{eqlinearized}
\end{equation}

The characteristic polynomial of (\ref{eqlinearized}), which determines all
solutions of (\ref{eqlinearized}) of the form
$\mathbf{v}(t)=\mathbf{v_0}e^{\lambda t}$, is  
\begin{equation}
(\lambda+b_1)(\lambda+b_2)(\lambda+b_3)+ de^{-\lambda \tau}=0,\ \ d=-f'(T_0) g_1 g_2 > 0   \label{eqchar}
\end{equation}

\begin{Prop}  The linear system (\ref{eqlinearized}) is stable, for all values
of $b_1,b_2,b_3,g_1,g_2,\tau$ and $f(x)=A/(K+x)$.      \label{teo:stability} 
\end{Prop}

\paragraph{Proof:} For there to be an unstable solution of (\ref{eqlinearized}), there must be a solution $\lambda$ of (\ref{eqchar}) such that $\mbox{Re } \lambda \geq 0$.  Assuming that this is the case, we have
\begin{equation}
d\geq |-d e^{-\lambda \tau}|=|\lambda+b_1||\lambda+b_2||\lambda+b_3| \geq
|b_1||b_2||b_3|=b_1b_2b_3\,.
\end{equation}
But on the other hand, using the choice for $f(T)$ above, we have
$f'(T_0)=-A/(K+T_0)^2=-f(T_0)/(K+T)$, and  
\begin{equation}
d=-f'(T_0) g_1g_2=\frac{f(T_0)}{K+T_0}g_1g_2=b_1b_2b_3\frac{T_0}{K+T_0}<b_1b_2b_3\,,
\end{equation}
which is a contradiction.  Q.E.D

\section{Global Asymptotic Stability of the Model}

Even with the addition of only one simple delay, it is probably best to view
(\ref{eqmodel}) as a dynamical system with states in the space $X$ of
continuous functions from $[-\tau,0]$ into the closed positive quadrant
$\Rplus^3$.  The right hand side of (\ref{eqmodel}) defines a function
$F:X\rightarrow \Rplus^3$ in the natural way, and given an initial state
$\phi\in X$, the solution of the system is the unique absolutely continuous
function $x:[-\tau,\infty)\rightarrow \Rplus^3$ such that
\begin{equation}
x_{(0)}=\phi \mbox{ and } \dot x(t)=F(x_{(t)}),\ t\geq 0  \label{closedloop} 
\end{equation}
Here, $x_{(t)}$, or simply $x_t$, is the state 
$\gamma(s)=x(t+s),\ s\in[-\tau,0]$.  The function $\Dyn(t,\phi)=x_t$ will be
from now on formally identified with system (\ref{eqmodel}).  For proofs of
the fact that $\Dyn$ is well-defined, and more details, the reader is referred
to \cite{Bens 1992,Hale 1993,Smith monotone}.

\paragraph{Cutting the Loop} We define a funcion $G:X\times \Rplus \rightarrow
\Rplus^3$ in a very similar manner to $F$: for $\phi(s)=(R(s),L(s),T(s))$, let
\[
F(\phi,w)=(w-b_1 R(0),\ g_1 R(0)-b_2 L(0),\ g_2L(t-\tau)-b_3 T(0)) \,.
\]
Given a piecewise continuous function $u:\Rplus\rightarrow \Rplus$, called an
{\em input}\footnote{We won't require the more general control--theoretic
definition where $u$ is measurable and locally bounded, see~\cite{Sontag
mono}}, we define $\Psi(t,\phi,u)=x_t$, where $x:[-\tau,\infty)\rightarrow
\Rplus^3$ is the unique absolutely continuous function such that  
\begin{equation}
x_{(0)}=\phi \mbox{ and } \dot x(t)=G(x_t,u(t)),\ t\geq 0. \label{openloop}
\end{equation}  
In effect, we are thus cutting the feedback loop induced by $T$ upon $R$, and
replacing it with an arbitrary input $u(t)$. 

Notation:  given $x,y\in \R^3$, let $x\leq y$ denote $x_i\leq y_i,\ i=1,2,3$.
For $\phi, \psi \in X$, let $\phi \leq \psi$ denote $\phi(s) \leq \psi(s),\
\forall s \in [-\tau, 0]$.

\begin{Teo}  \label{teo:properties}

The dynamical system with input $\Psi(t,\phi,u)$ satisfies the following
properties: 

\begin{enumerate}
\item If the input $u(t)$ converges to $w\in \Rplus$, then $\Psi(t,\phi,u)$
converges as $t$ tends to $\infty$ towards the constant state
\[
k(w)\;=\;
\left(\frac{w}{b_1},\frac{g_1w}{b_2b_1},\frac{g_1g_2w}{b_1b_2b_3}\right),
\]
for any initial state $\phi \in X$. 
\item Let $u_1,u_2$ be inputs, and pick any two initial states
$\phi,\psi \in X$. If \ $u_1(t)\leq u_2(t) \
\forall t$ and $\phi\leq \psi $, then $\Psi(t,\phi,u_1)\leq \Psi(t,\psi,u_2)
\ \forall t$.   
\end{enumerate}
\end{Teo}

\textbf{Proof:}  Suppose that $u(t)$ converges towards $w\in \Rplus$, and let
$\phi\in X$ arbitrary.  The dynamics of the component $R(t)$ of the solution
$x(t)$ is determined by the equation $\dot R(t)=u(t)-b_1R(t)$, and so $R(t)$ converges towards $w/b_1$.  Applying a very similar argument to $L(t)$ and $T(t)$ in this order, we obtain the first result.  

The proof of the second statement follows by the ``Kamke condition''
(see~\cite{Smith monotone}):  if $w_1\leq w_2,\ \phi\leq \psi$,  and
$\phi(0)_i=\psi(0)_i$ (that is, the $i$th components of $\phi$ and $\psi$ are
equal), then $G(\phi,w_1)_i\leq G(\psi,w_2)_i$.   For instance, if
$\phi=(R_1,L_1,T_1),\ \psi=(R_2,L_2,T_2)$, $\phi\leq \psi$, and
$R_1(0)=R_2(0)$, then $w_1-b_1R_1(0)\leq w_2-b_1R_2(0)$.  This can be checked
for $L$ and $T$ in the same way.   
The fact that the Kamke condition implies the desired property follows from
the results in~\cite{Smith monotone}; however, in the interest of exposition
and since the proof is so short, we provide it next.

Let $x(t)$ be the solution of (\ref{openloop}) with input $u_1$ and initial
condition $\phi$, and let $G_\epsilon=G + (\epsilon,\epsilon,\epsilon)$, for
$\epsilon >0$.  Let $y_\epsilon(t)$ be the solution of
$\dot{y}(t)=G_\epsilon(y_t,u_2)$ with initial condition $\psi$.  Suppose by
contradiction that at some point $t_1$, $x(t_1)\not\leq y_\epsilon(t_1)$, and
so there exists a component $i$ (that is, $R,L$ or $T$) and $t_0$ such that
$x_{t_0}\leq y_{\epsilon t_0},\ x(t_0)_i=y_\epsilon(t_0)_i$ and 
$\dot x(t_0)_i\geq \dot y_\epsilon(t_0)_i$.  But then 
\[
\dot x(t_0)_i = G(x_{t_0},u_1(t_0))_i\leq G(y_{\epsilon t_0},u_2(t_0))_i <
G_\epsilon(y_{\epsilon t_0},u_2(t_0))_i = \dot y_\epsilon(t_0)_i
\]
which is a contradiction.  We thus conclude that $x(t)\leq y_\epsilon(t),\ \forall t\geq 0$.  Now, it can be shown (\cite{Hale 1993},\cite{Smith monotone})  that as $\epsilon \rightarrow 0$ $y_\epsilon(t)$ converges pointwise to $y(t)$, the solution of (\ref{openloop}) with input $u_2$ and initial condition $\psi$, and from here the conclusion follows.  Q.E.D.

\vspace{2ex}

\begin{Def} Given $x:[-\tau,\infty)\rightarrow \Rplus^3$ be an arbitrary
trajectory, we say that $z\in \Rplus^3$ is a {\em lower hyperbound of $x(t)$}
if there is $ z_1,z_2, \ldots \rightarrow z$ and $t_1< t_2< t_3 \ldots
\rightarrow \infty$ such that for all $t\geq t_i,\ z_i \leq x(t)$.   A similar
definition is given if for all $t\geq t_i, \ z_i\geq x(t)$, and we say that
$z$ is an {\em upper hyperbound of $x(t)$}.   
\end{Def}
For instance, $z$ is a lower hyperbound of the trajectory $x$ if it bounds
from below $x(t)$ for every $t$.  Similar definitions are given for inputs
$u(t)$.  The previous Theorem is the basis for the following result.  

\begin{Teo}   \label{teo:underbound}
Let $v\in \Rplus$  be a lower hyperbound of the input $u(t)$, and let 
$\phi\in X$ be arbitrary.  Then $k(v)$ is a lower hyperbound of the solution
$x(t)$ of the system (\ref{openloop}).  If $v$ is, instead, an upper
hyperbound of $u(t)$, then $k(v)$ is an upper hyperbound of $x(t)$.       
\end{Teo}

\paragraph{Proof:}  Suppose that $v$ is a lower hyperbound of $u(t)$, the
other case being similar, and let $v_1, v_2, \ldots \rightarrow v$ and 
$t_1<t_2 < \ldots \rightarrow \infty$ be as above.  

For every $i\geq 1$, let $y_i\in \Rplus^3$ and $V_i \subset \Rplus^3$
neighborhood of $k(v_i)$ that is open in $\Rplus^3$, chosen in such a way that
$y_i \leq V_i$ and $|y_i-k(v_i)|\leq 1/i$.  Also, let
\[
u_i(t)=\left\{ \begin{array}{ll}  u(t), & 0\leq t < t_n \\
                        v_n,  & t \geq t_n \, .   \end{array} \right.
\]
Let $T_1 < T_2 < \ldots \infty$ be defined by induction as follows:  $T_1=0$, and if $T_{i-1}$ is defined, let $T_i$ be chosen such that $T_i \geq T_{i-1},\ T_i \geq t_i$, and for all $t\geq T_i:\ x_i(t)=\Psi(t,\phi,u_i)\in V_i$.  By the previous theorem, $x_i(t) \leq x(t) \forall t$, and so $y_i \leq x(t),\ \forall t\geq T_i$.  As $y_i \rightarrow k(v)$, the conclusion follows.  Q.E.D. 

\vspace{2ex}

The following simple Lemma is standard in the literature on discrete
iterations (and is used in a similar context in~\cite{predator prey});
we provide a proof for expository purposes.

\begin{Lema}
Let $S:\Rplus\rightarrow \Rplus$ be a continuous, nonincreasing function.  Then the discrete system $u_{n+1}=S(u_n)$ has a unique, globally attractive equilibrium if and only if the equation $S(S(x))=x$ has a unique solution. 
\end{Lema}

\paragraph{Proof:}  If the system has a unique, globally attractive
equilibrium $\bar{u}$, then this point is a solution of the equation
$S^2(x)=S(S(x))=x$.  Any other point $u$ cannot be a solution of this
equation, as $S^n(u)$ must converge to $\bar{u}$.  This proves
one of the directions of the lemma.   

Conversely, suppose that the equation $S^2(x)=x$ has a unique solution.  Let
$u\in \Rplus$ be arbitrary, and consider the sequence
$u_n=S^n(u)$.  If $u\leq u_2$, then
since $S^2$ is a nondecreasing function, we have $u_2 \leq u_4$, and so
\[
u \leq u_2 \leq  u_4 \leq u_6 \leq \ldots \,.
\]
But the sequence $u_2,u_4 \ldots$ is bounded (by $S(0)$), 
and so $u_{2n}$ must converge to some point $v_0$.
The same argument applies if $u_2< u$, and also for the sequence
$u_1,u_3,u_5,\ldots$, which must converge to some point $v_1$.   But the
continuity of $S$ implies that both $v_0$ and $v_1$ are solutions of
$S^2(x)=x$, so $v_0=v_1$ are both equal to our unique solution, and $u^n$
thus converges to this point, independently of the choice of $u$.  Q.E.D.   

\vspace{2ex}

Consider for instance $S(x)=p/(q+x)$, where $p,q$ are positive real numbers.  If $x$ satisfies $S^2(x)=x$, then it holds that 
\[
 x=\frac{p}{q+S(x)}
\]
which can be rearranged as $x^2+qx-p=0$.  Using the quadratic formula, it becomes clear that there is always exactly one {\em positive} solution.

This example will be useful in what follows.  

\begin{Teo} All solutions of the system (\ref{closedloop}), with $f=A/(K+x)$,
converge towards the unique equilibrium, for any choice of the parameters
$b_1,b_2,b_3,g_1,g_2,\tau,A,K$.
\end{Teo}

\paragraph{Proof:}  Consider any initial condition $\phi\in X$, and the
corresponding solution $x(t)=(R(t),L(t),T(t))$ of (\ref{closedloop}.
Defining the input $u(t)=f(T(t))$, and using it to solve the system
(\ref{openloop}) with initial condition $\phi$, we arrive of course at exactly
the same solution $x(t)$.    

Let $v$ bound $u(t)$ from below for all $t$ -- for instance, $v=0$ will do.
Then by Theorem~\ref{teo:underbound}, $k(v)$ is a lower hyperbound of $x(t)$.
In particular,
\[
Qv=\frac{g_1g_2}{b_1b_2b_3}v
\]
is a lower hyperbound of $T(t)$.
But, since $f$ is a nonincreasing function,
this implies that $f(Qv)$ is an upper hyperbound of $f(T(t))=u(t)$.
Defining $v_1=f(Qv)$, we apply the same theorem once again to show that
$k(v_1)$ is an upper hyperbound of $x(t)$, $v_2=f(Qv_1)$ is a lower hyperbound
of $u(t)$,  etc.  But  
\[
f(Qx)=\frac{A}{K+Qx}=\frac{p}{q+x}=S(x)
\]
for $p=A/Q,\ q=K/Q$.  Thus we see that $v_n=S^n(v)$ is a convergent sequence
of numbers that are alternatively upper and lower hyperbounds of $u(t)$.  This
easily implies that $u(t)$ itself is a trajectory that converges to the unique
solution $\bar{u}$ of the equation $S^2(x)=x$.  By
Theorem~\ref{teo:properties}, $x(t)$ converges towards $k(\bar{u})$,
independently of the choice of the initial condition $\phi$.   

Finally, this implies that $k(\bar{u})$ is the unique equilibrium of the
system, otherwise one could reach a contradiction by taking this equilibrium
as constant initial condition.  Q.E.D. 

\vspace{2ex}

\subsection{Discussion}

Several remarks are in order:  first, the value of the
delay $\tau$ was almost never used,
and indeed can be arbitrarily large or small.  In fact, we can introduce
different delays, large or small, in all of the first summands of the right
hand sides of~(\ref{closedloop}), and the results will apply with almost no
variation.  If delays are introduced in the second summands, the system will
not be {\em monotone}, that is, won't satisfy the second property of
Theorem~\ref{teo:properties}, which is essential for this argument.  But then
again, introducing a delay in the degradation terms wouldn't be very
biologically meaningful.  For more on monotone systems, the reader is referred
to the excellent textbook by Hal Smith~\cite{Smith monotone}, and \cite{Hirsch 1989}.

The above argument is an illustration of a more general treatment on a class
of delayed dynamical systems with monotone subsystems and negative feedback
interconnection. The underlying  order may be generalized as  $x\leq y$ iff
$y-x$ lies in a {\em cone} $K\subseteq \R^3$ (see~\cite{Sontag mono}).  This
provides for more generality and applicability in biological problems.  The
key sufficient condition is that the discrete dynamical system
$u_{n+1}=S(u_n)$ be globally attractive; in a sense the dynamics of the
continuous system is reduced to that of the discrete one, which may eventually
involve state spaces with substantially fewer dimensions.  
See~\cite{Sontag mono},\cite{Sontag multi} and work to appear by the present
authors for this more general treatment.    

As for the conclusions in pp. 244-253 of {\em Mathematical Biology}, we may venture to suggest that in eq 7.49, p. 247, the author writes the characteristic equation (\ref{eqchar}) of the linearized system~(\ref{eqlinearized}) as 
\begin{equation}
\lambda^3 + a \lambda^2 + b\lambda + c + d e^{-\lambda \tau}=0
\end{equation}
where $a,b,c,d$ are all written in terms of the original parameters of the system: $a=b_1+b_2+b_3$, etc.  From here on the efforts are concentrated in finding a root $\lambda$ of this equation with $\mbox{Re } \lambda = 0$, for some well-chosen coefficients $a,b,c,d$.  But the author seems to disregard in the remaining argument the fact that $a,b,c,d$ cannot be chosen arbitrarily and independently, {\em but rather that their values are determined from choosing arbitrarily $b_1,b_2,b_3,g_1,g_2,\tau$}.  Thus for instance, it is assumed in the last line of p.251 that $d>c$, without justification from the original variables.  The former assumption turns out not to be possible to satisfy for the particular choice of $f$, as seen in the proof of Proposition~\ref{teo:stability}.     
 
We point out that a simple modification can make oscillatory behavior
possible.  In p. 246 of \cite{Murray 2002}, the author discusses varying
cooperativity coefficients of $f(x)=A/(K+x^m)$, then settles for $m=1$ for the
delayed model.  If indeed $m$ is increased, then it is very possible to have
$d>c$ and the remaining argument in the section will be valid.  One example of
this is when parameters are picked as follows:
\[
m=2,\, A=10,\, K=2,\, b_1=1,\, b_2=1,\, b_3=1,\, g_1=10,\, g_2=10
\, .
\]

Another interesting contribution to the modeling of testosterone dynamics is
the paper~\cite{Ruan 2001} by Ruan et al., where sufficient conditions are
found for stable 
and oscillatory behavior in a neighborhood of an equilibrium.  We would like
to describe the relationship between~\cite{Ruan 2001} and our
own result, given the similarity of the hypotheses and the potentially
conflicting conclusions: global stability in our results vs.\ Hopf
bifurcations in~\cite{Ruan 2001}.  
Moreover, we will simplify the statement of that result.
In that paper, several new quantities are introduced in order to state the main
result, Theorem~3.1.  In terms of the original variables of the system
($b_1$, $b_2$, $b_3$, etc.), these are as follows:
\begin{eqnarray*}
p&=&b_1^2+b_2^2+b_3^2 \geq 0\\
q&=&b_1^2b_2^2 + b_2^2b_3^2 + b_1^2b_3^2 \geq 0\\
\Delta&=&p^2-3q=\frac{1}{2}((b_1^2-b_2^2)^2+ (b_2^2-b_3^2)^2+ (b_1^2-b_3^2)^2) \geq 0\\
 z_1&=&\frac{1}{3}(-p+\sqrt{\Delta})  \,.
\end{eqnarray*}
Theorem 3.1 holds under the assumption that
\begin{equation}
\label{bdcond}
(b_1 + b_2)(b_1 + b_3)(b_3 + b_2)   < d
\end{equation}
and deals essentially with three following three special cases:
\begin{enumerate}
\item 
$b_1b_2b_3 \geq d$
and $\Delta < 0$,  
\item 
$b_1b_2b_3 \geq d$ and
$z_1>0$,
\item
$b_1b_2b_3 < d$ \,.
\end{enumerate}
In case 1, (local) asymptotic stability is guaranteed for arbitrary delay
lengths (part (i) of the Theorem), while
in cases 2 and 3, and under some additional conditions
(parts (ii) and (iii) of the Theorem), stability holds for
small enough delays, but a Hopf bifurcation occurs at some critical value of
this delay length.
In light of the above computation, case 1 can never be satisfied
(for variables $p$, $q$, $r$ generated from the original set of parameters
$b_1$, $b_2$, $b_3$, etc.). 
Similarly, the condition $z_1>0$ will never be satisfied, since
$$ z_1>0 \Leftrightarrow  \Delta> p^2  \Leftrightarrow 3q < 0 $$
so case 2 cannot hold either.
One is only left with case 3, which is actually a consequence
of~(\ref{bdcond}).
On the other hand, for the particular choice of $f(x)$ made in \cite{Murray
2002} and the present 
paper, Proposition~\ref{teo:stability} shows that we always have
$b_1b_2b_3>d$.  Thus Theorem 3.1 does not apply for the present model, as well
as for any choice of the function $f$ and any set of parameters such that
$b_1b_2b_3>d$.

\paragraph{Acknowledgement:} We would like to thank Augusto Ponce for
useful suggestions.

\end{document}